\documentclass[aps,prl,twocolumn,groupedaddress]{revtex4}
\usepackage{color,amsmath, amssymb, graphicx, amsfonts , setspace}

\newcommand{\ua}{\ensuremath{\uparrow}}
\newcommand{\da}{\ensuremath{\downarrow}}

\newcommand{\dg}[0]{\ensuremath{\dagger}}
\newcommand{\av}[1]{\ensuremath{\left\langle #1 \right\rangle}}
\newcommand{\dif}{\ensuremath{\,\mathrm{d}}} 

\begin{document}


\title{Signature of the FFLO phase in the collective modes of a
trapped ultracold Fermi gas}


\author{Jonathan M. Edge}
\affiliation{T.C.M. Group, Cavendish Laboratory, J.~J.~Thomson Ave., Cambridge CB3~0HE, UK.}
\author{N. R. Cooper}
\affiliation{T.C.M. Group, Cavendish Laboratory, J.~J.~Thomson Ave., Cambridge CB3~0HE, UK.}


\date{\today}

\begin{abstract}
%
We study theoretically the collective modes of a two-component Fermi
gas with attractive interactions in a quasi-one-dimensional harmonic
trap. We focus on an imbalanced gas in the
Fulde-Ferrell-Larkin-Ovchinnikov (FFLO) phase. Using a mean-field
theory, we study the response of the ground state to time-dependent
potentials.  For potentials with short wavelengths, we find dramatic
signatures in the large-scale response of the gas which are
characteristic of the FFLO phase. This response provides an effective
way to detect the FFLO state in experiments.
\end{abstract}


\maketitle


With the advent of ultracold atomic gases new states of matter are
becoming experimentally accessible. Important developments have
included the achievement of superfluidity in two-component Fermi
gases \cite{bloch07}, and the study of the effects of
density imbalance in these
systems\cite{pairing_and_phase_sep_in_pol_fg_partridge_hulet}.
These developments open up the prospect of the study of unconventional superfluid
phases, in particular the inhomogeneous superfluid state of
Fulde-Ferrell-Larkin-Ovchinnikov (FFLO)
\cite{superconductivity_in_spin_exch_field_fulde_ferrell64}.
This state has been under
theoretical investigation since the 1960s, playing a central role in
our understanding of superfluidity of fermions both within condensed
matter physics and in particle physics \cite{inhomogeneous_sc_in_cond_mat_and_qcd-casalbuoni_nadulli}; yet to date no
unambiguous demonstration of the FFLO state has been achieved.

While theory predicts the appearance of the FFLO phase in a bulk (3D)
gas of ultracold fermions  \cite{bec-bcs_xover_in_pol_res_SF-radzihovsky} the region of
parameter space that it occupies is expected to be very small. It has
been shown that this phase is greatly stabilised in (quasi-)1D
geometries \cite{orso_attr_fermi_gases_bethe_ansatz07,phase_diag_of_str_int_pol_fg_1d_drummond07,various_1d_studies,fflo_state_in_1d_attr_hubbard_model_luescher_laeuchli08}. 
Such
geometries can readily be achieved in atomic gases by optical
confinement\cite{ex_coll_osc_in_1d_gas_moritz_esslinger03}, so
experiments on ultracold Fermi gases will soon be in the regime where
the FFLO phase can be studied.
In view of this, it is important and timely to ask: what are the
observable properties of the FFLO phase?  

Existing proposals for the detection of the FFLO phase include the
probing of noise correlations in an expansion image following release
of the trap\cite{fflo_state_in_1d_attr_hubbard_model_luescher_laeuchli08}, and the use of RF spectroscopy to excite atoms
out of the gas into other states
\cite{spectral_signatures_of_fflo_in_1d_bakhtiari_torma}.  In
this Letter we show that characteristic signatures of the spatial
inhomogeneity of the FFLO phase can be found in the collective mode
response of the trapped gas. The method directly probes the 
intrinsic inhomogeneity of the FFLO phase -- by the use of a
perturbation by an
potential with a short-wavelength
spatial periodicity. An important feature of this method is that the
signatures of the microscopic nature of the phase appear in the
response on very {\it large} length scales (the scale of the atomic
cloud). Consequently, the measurements of the response do not require
high spatial resolution, and could be performed in situ without
expansion of the gas.

We study a model of two-component fermions in 1D with attractive
contact interactions, and
unequal densities.
For a homogeneous system, the exact groundstates are known from the
Bethe ansatz \cite{un_systeme_a_une_dimension_de_fermions_en_interaction-gaudin67,phase_diag_of_str_int_pol_fg_1d_drummond07,orso_attr_fermi_gases_bethe_ansatz07}. Depending on the densities of the two
components there are three phases that appear at $T=0$: the
unpolarised superfluid state; a fully polarised (therefore
non-interacting) Fermi gas; and a partially polarised phase \cite{orso_attr_fermi_gases_bethe_ansatz07}, associated with the FFLO state.
This is a state that, within mean-field theory, has a local
superconducting gap that oscillates in space.
As has been discussed in detail by Liu et al.\cite{liu_drummond_07},
mean-field theory provides an accurate description of the exact phase
diagram for the 1D system, at least within the weak-coupling BCS
regime.
We have extended the mean-field theory to investigate the linear
response and collective modes of both the trapped and untrapped
imbalanced Fermi gas. As we shall discuss in detail later, this approximate
theory correctly captures all qualitative features of the collective
modes that are important for our purposes.  

To effect numerical calculations, we study a discretized version of
the problem (the attractive Hubbard model), for which the mean-field
Bogoliubov-de Gennes (BdG) Hamiltonian is
$
\hat H  =   -J\sum_{i,\sigma}\left( \hat c^\dg_{i+1, \sigma} \hat c_{i, \sigma} + h.c. \right)
+ \sum_i \left( \Delta_i \hat c^{\dg}_{i, \ua} \hat c^{\dg}_{i, \da} + h.c.\right)
+ \sum_{i, \sigma} W_{i, \sigma} \hat c^{\dg}_{i, \sigma} \hat c_{i, \sigma}
$
where $\hat c_{i,\sigma}^{(\dg)}$ are fermionic operators for species $\sigma = \ua, \da$ on site $i$,
$W_{i, \sigma} \equiv V^{ext}_i + U
\langle \hat c^\dag_{i,\bar\sigma} \hat c_{i,\bar\sigma}\rangle - \mu_\sigma$ (with $V^{ext}$ the external potential and
$\mu_\sigma$ the chemical potentials), and $\Delta_i \equiv U
\av{\hat c_{i, \da} \hat c_{i,\ua}}$ is the local superfluid gap. $J$ is the
hopping parameter and $U$ the on-site interaction strength ($U<0$ is
assumed throughout). 
We derive the linear response
to external time-dependent perturbations, $\delta W_{i,\sigma}(t)$, by
supplementing the (self-consistent) solutions of the above BdG equations with the random
phase approximation (RPA) \cite{bruun_mottelson01}. Divergences in the
response appear at the frequencies of the collective modes of the
system.
The results we present are at sufficiently low
particle density to be representative of the continuum limit.  In the
mapping to the continuum, we relate site $i$ to position $x$ via
$x=ia$ and the mass is $m = \frac{\hbar^2}{2 J a^2}$.
The interaction strength $\gamma$, defined as the ratio of the interaction energy density to the kinetic energy density, is given by $\gamma = -\frac{m g_{1d}}{\hbar^2\rho}$ \cite{liu_drummond_07} where $g_{1d} = \frac Ua$, $\rho$ is the total density of particles and $E_F \equiv \pi^2\rho^2 J/4$.

We discuss first the response of a homogeneous system, with
$V^{ext}=0$ and periodic boundary conditions. We focus on the FFLO
phase, with unequal average particle densities, for which the
self-consistent mean-field solution has an oscillating gap $\Delta(x)$
with wavelength $\lambda_\Delta = 2\pi/(k_{F,\ua}-k_{F,\downarrow})$,
where $k_{F,\sigma}$ are the Fermi wavevectors for the non-interacting
imbalanced gas. We study the response to periodic perturbing potentials
 \begin{equation}
\delta W_{\sigma}(x,t) = V_{\sigma} \sin kx \cos\omega t
\label{eq:vlambda}
\end{equation}
We refer to the case $V_\ua = V_\da = V_0$ as ``spin-symmetric'' and
$V_\ua = -V_\da = V_0$ as ``spin-asymmetric'' excitation.  
\begin{figure}
  \centering
  \includegraphics[width = 8.5cm]{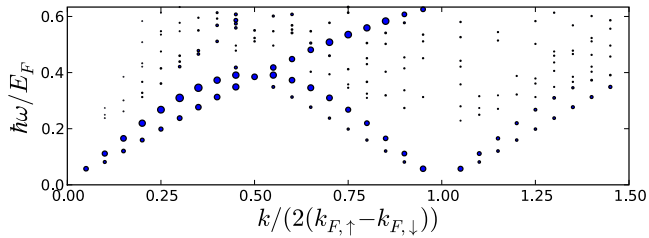}
  \caption{Response
    \protect\cite{comment_in_fig_1}
    of a homogeneous system in the FFLO phase to a
    spin-asymmetric periodic potential (\protect\ref{eq:vlambda}).
    The area of the circle is proportional to the amplitude of the response.
    The
  polarisation $p \equiv \frac{k_{F\ua} - k_{F\da}}{k_{F\ua} + k_{F\da}} = 0.15$, and the
  interactions strength is $\gamma = 1.5$.  Two gapless sound modes are seen
  to emerge around $k\simeq 0$ and $k\simeq k^* \equiv 2
  (k_{F_\ua}-k_{F\da})$. The simulation was done on 270 lattice sites.
  }
  \label{fig:two_sound_modes}
\end{figure}

The response of an imbalanced gas in the FFLO phase to a
spin-asymmetric modulation is shown in Fig.~\ref{fig:two_sound_modes}.
At low frequency and long wavelength there appear two distinct sound
modes with different velocities \cite{frahmvekua,spectral_props_of_part_pol_1d_luttinger_sf_feiguin_huse09}.  
 Within 
mean-field theory, the appearance of these two gapless modes arises
from the fact that the FFLO phase breaks both gauge symmetry and
translational symmetry. (We later discuss these properties beyond mean field theory.)
An analysis of the two low-frequency modes
shows that they are of mixed character, involving spatial oscillations
of both density and spin-density. This is expected from the breaking
of time reversal symmetry by the imbalance\cite{frahmvekua}, and is
consistent with the results (in Fig.~\ref{fig:two_sound_modes}) that both modes can be excited
by a purely spin-asymmetric perturbation, $V_\uparrow=-V_\downarrow$.

In Fig.~\ref{fig:two_sound_modes} it is clear that linear sound modes
emerge also around the point $k = k^*\equiv 2(k_{F\ua} - k_{F\da})$.  Within mean-field theory, this too can
be understood as a consequence of the broken translational symmetry of
the FFLO phase. This leads to a Brillouin zone structure for the
collective modes, characterised by a reciprocal lattice vector of size
$k^*$.  Note that this is twice the value that one would expect from
the translational periodicity of the gap, $\lambda_\Delta$.
To
understand this, observe that, while the gap $\Delta$
has period $\lambda_\Delta$, the densities
$\rho_\ua(x) $ and $\rho_\da(x)$ have periods
$\lambda_\Delta/2$, and the mean-field ground state has an exact symmetry 
under the transformation
$
  x \to x + \lambda_\Delta/2
  ,\;
  \Delta \to -\Delta
$.
Applying this symmetry to the RPA response calculation leads to a generalised
Bloch theorem for the collective modes. The wavevector $k$ is found to
be conserved modulo $(2\pi)/(\lambda_\Delta/2) = k^*$, consistent with
the response in Fig.~\ref{fig:two_sound_modes}. 
For a spin-symmetric perturbation a similar qualitative response is
found. However, the response close to $k=k^*$ is smaller than for the
spin-asymmetric perturbation.  
We account for this by the fact that the
periodic density of excess majority particles manifests itself more
strongly in the difference of the two densities than in the sum of the
two densities.  We now turn to discuss the manifestations of this response for
a trapped gas.

\begin{figure}
  \centering
  \includegraphics[width=4.25cm]{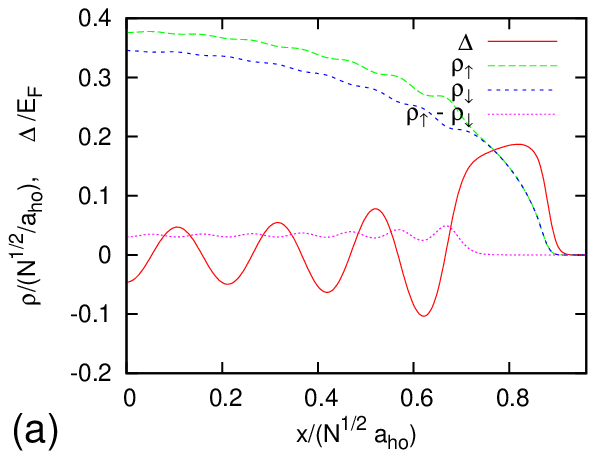}
  \hspace{0.1cm}
  \includegraphics[width=4.cm]{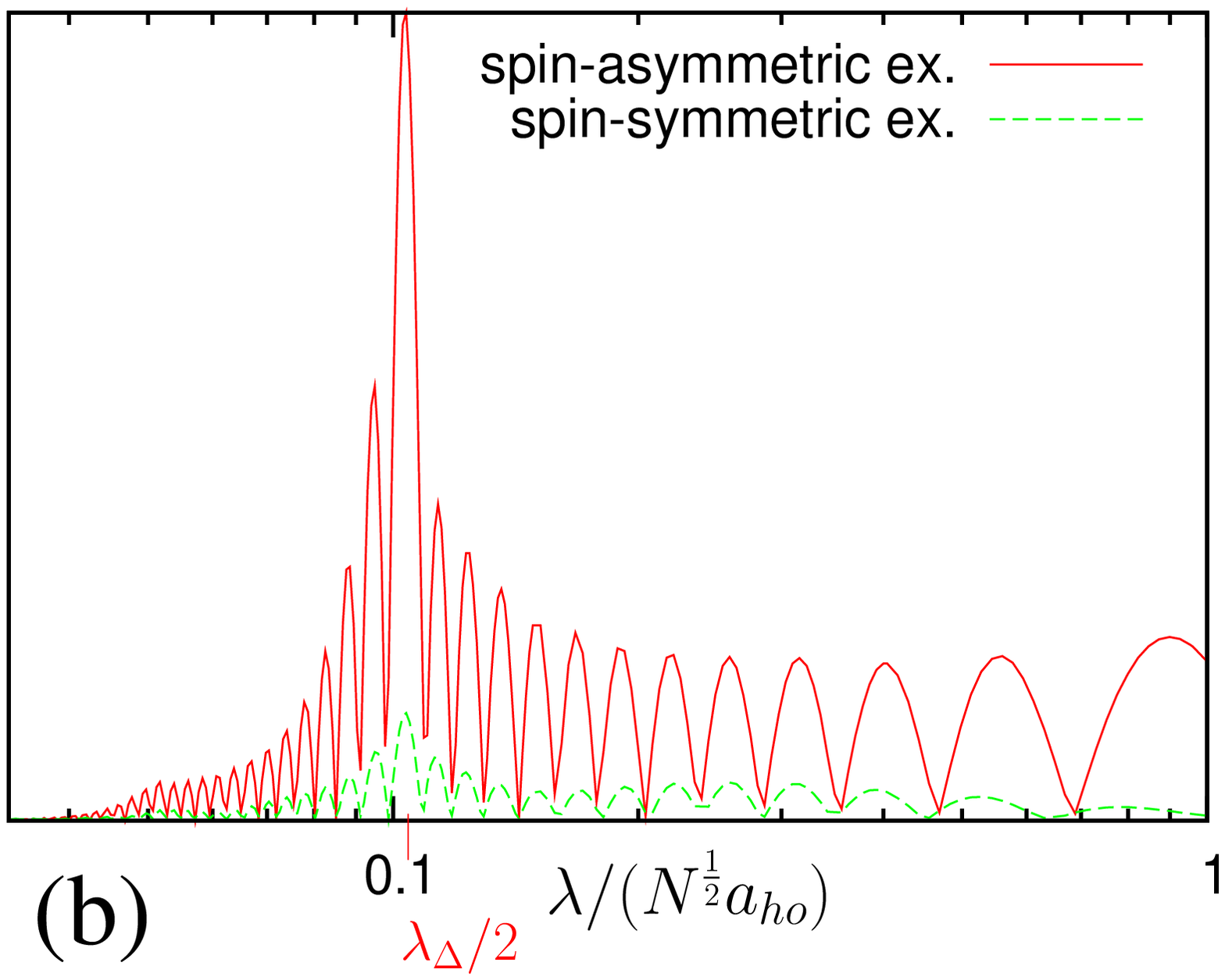}
  \caption{(Colour online.) Configuration with fully paired phase towards the edge of the
    trap. (a): densities and the value of the superfluid gap
    $\Delta$. (b): response of the spin-dipole mode to
    excitations of different wavelengths (arbitrary units). Here $p=0.048$, $\gamma = 0.93$ (measured in centre), number of particles $N = 290$, lattice spacing $a = 3.3\cdot10^{-3} N^{\frac12}a_{ho} $. 
    The perturbing potential
    has a fixed amplitude, while the wavelength is varied. }
  \label{fig:full_paired_towards_edge}
\end{figure}

\begin{figure}
  \centering
  \includegraphics[width=4.25cm]{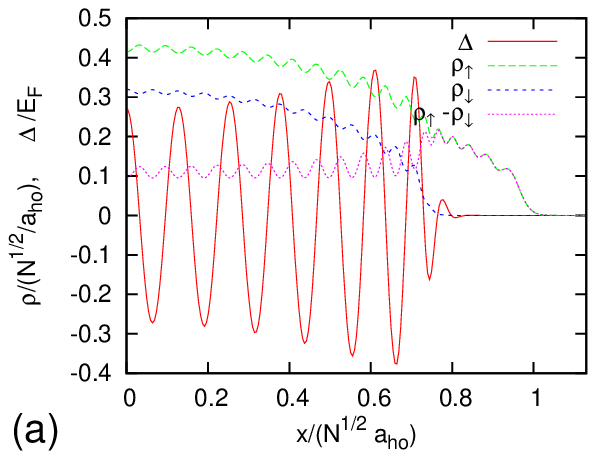}
  \hspace{.1cm}
  \includegraphics[width=4.cm]{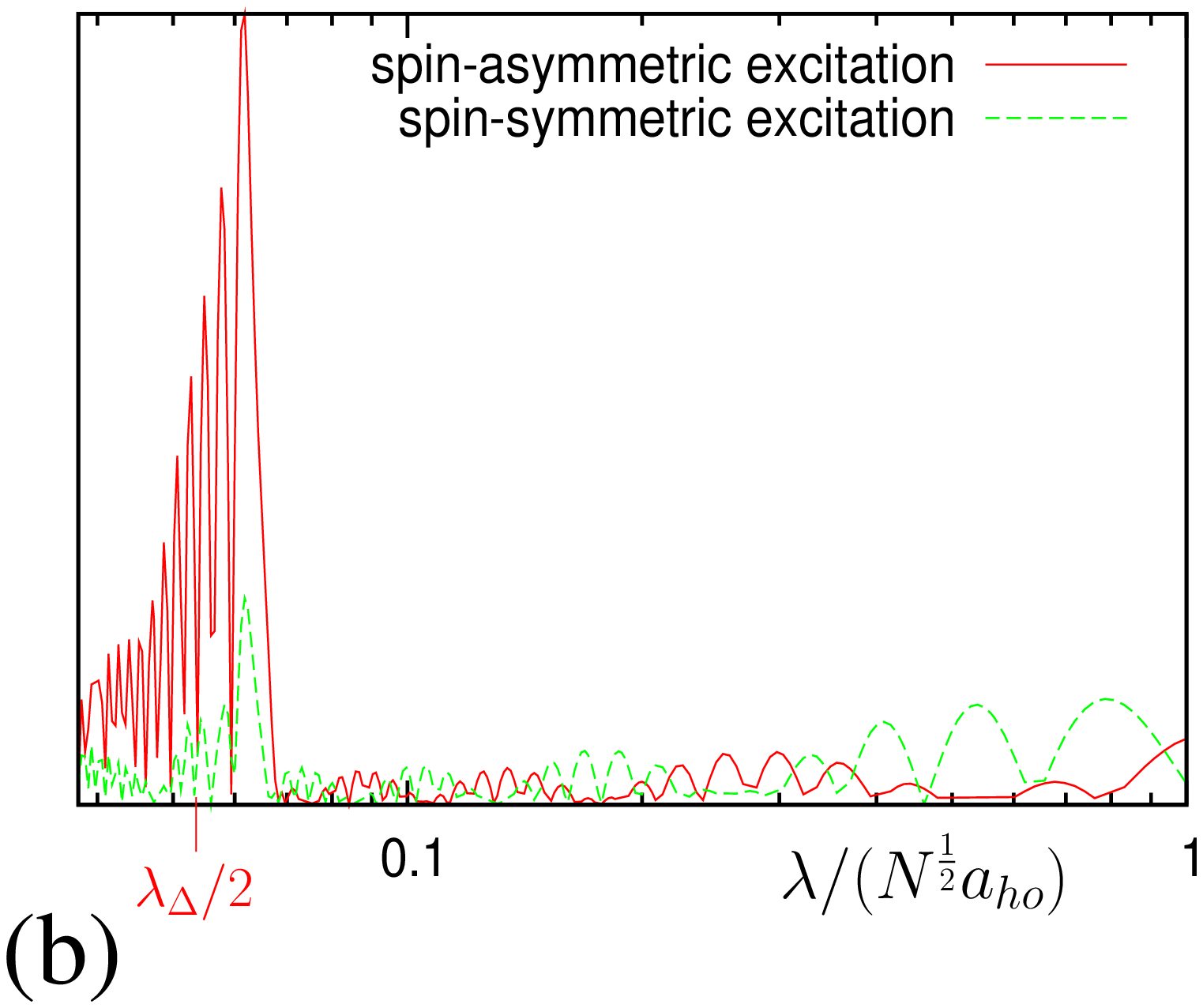}
  \caption{(Colour online.) Configuration with fully polarised phase towards the edge. Here $p=0.25$, $\gamma = 1.5$, $a/(N^{\frac12}a_{ho}) = 4.7\cdot10^{-3}$, $N=143$ . Otherwise the same as fig.~\ref{fig:full_paired_towards_edge}.}
  \label{fig:full_polarized_towards_edge}
\end{figure}

We have studied the attractive Fermi gas in a harmonic trap, with $V^{\rm
ext}(x) = \frac{1}{2} m \omega_0^2 x^2$ and with unequal particles
numbers $N_\uparrow\neq N_\downarrow$.  According to
\citet{orso_attr_fermi_gases_bethe_ansatz07} a 1D imbalanced trapped
Fermi gas can only be in one of two configurations: (a) A partially
polarised phase in the middle, with a fully paired phase towards the
edge; (b) a partially polarised phase in the middle, with a fully
polarised phase towards the edge.  These configurations appear within
the self-consistent BdG mean-field theory\cite{liu_drummond_07}. An
example of the case (a) is given in
Fig.~\ref{fig:full_paired_towards_edge}, and of the case (b) in
Fig.~\ref{fig:full_polarized_towards_edge}, where lengths are measured
in units of $N^{1/2} a_{\rm ho}$ (with $N = N_\ua +N_\da$ the total
number of fermions, and $a_{\rm ho} \equiv \sqrt{\hbar/m\omega_0}$ the
oscillator length of the trap). The configurations in Figs.~\ref{fig:full_paired_towards_edge} and \ref{fig:full_polarized_towards_edge} are in the weak coupling BCS regime, having values of $\gamma$ for which BdG gives qualitatively the right density profiles \cite{liu_drummond_07}.
We have studied the response of the trapped imbalanced Fermi gas in
both these regimes.

Perturbations with wavelength much longer than the size of the system
couple via potentials $V_\sigma(x)$ that are proportional to $x$.  We
find that the response to such potentials is well described by two
sharp modes, involving the dipolar oscillations of the density and
spin.  One of these modes -- the ``Kohn mode'' -- involves the
in-phase motion of both spin components, and has frequency $\omega =
\omega_0$, independent of the state of the system. This exact result
holds also within RPA\cite{Ohashi_on_Kohn_mode03}, and is recovered to
high accuracy in our calculations, showing that discretisation effects
are minimal.  The other mode -- the ``spin-dipole'' mode -- involves
the relative motion of the two spin species. For attractive
interactions the frequency lies slightly above $\omega_0$ (at
$\omega_{\rm sd} = 1.33\omega_0$ and $1.28\omega_0$ for
Figs.~\ref{fig:full_paired_towards_edge} and
\ref{fig:full_polarized_towards_edge} respectively).
While the frequency of the spin-dipole mode does depend on the state
of the system, it is does not show strong features of the presence of
the FFLO phase.  Similarly, the breathing modes are insensitive to
the microscopic nature of the atomic
gas\cite{phase_diag_of_str_int_pol_fg_1d_drummond07}.

The most interesting features arise when one excites the gas by time
dependent potentials (\ref{eq:vlambda}) with short wavelengths. We
find characteristic signatures of the microscopic nature of the trapped
gas in the response of spin-dipole mode -- i.e. on length scales much
larger than the wavelength of the applied
perturbation.\footnote{Qualitatively similar behaviour is found for
the Kohn mode, however with smaller amplitude.} In
Figs~\ref{fig:full_paired_towards_edge}(b) and
\ref{fig:full_polarized_towards_edge}(b) we show the amplitude of the
response of the spin-dipole mode to spin-symmetric and spin-asymmetric
periodic perturbations (\ref{eq:vlambda}) as a function of wavelength
$\lambda = 2\pi/k$.

For wavelengths $\lambda$ that are short compared to the size of the
cloud, $N^{1/2} a_{\rm ho}$, but large compared to $\lambda_\Delta$,
we find an oscillatory response as a function of $\lambda$ in both
Figs.~\ref{fig:full_paired_towards_edge}(b) and
\ref{fig:full_polarized_towards_edge}(b).  To understand this
behaviour, we note that the spin-dipole mode can be excited when the
two species experience a net difference in their acceleration.
This difference is
$  a(k) = \frac {V_0 k}{m} \int \left[\rho_\ua(x)/N_\ua \mp \rho_\da(x)/N_\da\right] \cos(kx) \dif x$
where the $\mp$ signs denote spin-symmetric/spin-asymmetric
excitations.   The difference in
acceleration oscillates with $k$, in a manner that is largely
controlled by the sizes of the clouds.
For case (b) we assume, for simplicity, parabolic density distributions of size $x_\sigma$ for the two spin components $\sigma$ \footnote{Similar results hold for other density distributions which fall off sharply towards the edge.}. Defining
$r \equiv \frac{x_\ua + x_\da}2$ and
$\delta\equiv \frac{x_\ua - x_\da}2$, the condition for a maximum is
$\sin kr \cos k\delta =0$ for spin-asymmetric excitations, and $\cos
kr\sin k\delta =0$ for spin-symmetric excitations. For
$\delta \ll r$ it follows that for spin-asymmetric (spin-symmetric)
potentials maxima in the response come from the $\sin kr$ ($\cos kr$)
term and the antinodes in the envelope come from the $\cos k\delta$
($\sin k\delta$) term.
Case (a) can be described by assuming a parabolic density distribution for the average density but where there is a constant difference between the two densities in the partially polarised region of size $r_p$. For spin-symmetric and spin-asymmetric excitations the condition for a peak is given by $\cos k r_p =0$.
The modulation of the peaks by this envelope structure allows us to
distinguish the two imbalanced trap configurations (a) for which the
two clouds have equal size $x_\ua = x_\da$ and there is no modulation and
(b) for which $x_\ua\neq x_\da$ and a modulation appears, see figs.~\ref{fig:full_paired_towards_edge}(b) and \ref{fig:full_polarized_towards_edge}(b).

As the wavelength is further reduced, we find dramatic signatures of
the presence of the FFLO phase, in both the cases (a) and (b)
described above.
Specifically, we find an unusually large response of the
amplitude of the spin-dipole mode when the wavelength of the
excitation becomes $\lambda = \frac{\lambda_\Delta}2$.
This response can be
understood in terms of coupling to the gapless modes at
$k=\frac{4\pi}{\lambda_\Delta} = k^*$ of the homogeneous system (see
Fig.\ref{fig:two_sound_modes}). A small deviation from $k^*$ (of the
order the inverse system size) allows mixing of this mode to the
spin-dipole mode, causing the response at $k^*$ to be apparent in the
dipolar motion of the atomic cloud.  

While the calculations we have presented are within RPA mean-field
theory, as we now argue, the qualitative conclusions are valid more
generally. Essentially, our results rely on the fact that in the
unpolarised FFLO phase, there exists low-frequency (of order
$\omega_0$) response that is sharply peaked at a spatial wavevector
$k^*$.  In the above, this response was accounted for in terms of the
broken translational symmetry of the mean-field state. As is well
known, in a true 1D quantum system no continuous symmetries are
broken. Thus, the broken (phase and translational) symmetries of the
mean-field FFLO state can lead only to power-law decay of the
respective correlation
functions\cite{fflo_pairing_in_1d_opt_latt-rizzi_fazio08}.  However,
the qualitative features described above are the same as those
expected from an exact treatment of the system. The transition from the
(unpolarised) superconducting phase to the partially polarised phase
is marked by the closing of the spin-gap, leading to a second gapless
sound mode. This can be viewed as a Luttinger liquid representing the
excess fermions\cite{Yang_inhomogeneous_sc_state_1d_01}.  These excess
particles have density $\rho_\ua-\rho_\da$, and thus a Fermi wavevector
$(k_{F_\ua} - k_{F_\da})$.  Thus, there are two gapless collective
modes, arising from the fully paired particles, and the liquid of
excess majority spin particles.  Furthermore, as in the general theory
of Luttinger liquids\cite{HaldaneJPhysC} the spectral function of
these excess majority spin particles will show gapless response at
multiples of twice their Fermi wavevector. This is $2(k_{F_\ua} -
k_{F_\da}) = 4\pi/\lambda_\Delta$ which is precisely the wavevector
$k^*$ at which one finds the gapless response in mean field
theory. 
 Thus, the qualitative features of the collective
excitation spectrum obtained in mean-field theory are fully consistent
with those expected for the exact system.

Any distinction between the exact results and those of mean-field
theory will be further reduced in the finite-size systems on which
experiments can be performed, where the distinction between long-range
and power-law correlations becomes blurred.  Indeed, one can expect
that {\it density inhomogeneity} will play a more significant role
than will the power-law decay of correlations. Since the densities of
spin-up and spin-down particles are inhomogeneous, so too is (the
local value of) $\lambda_\Delta$, so one can expect smearing of the
condition $\lambda = \lambda_\Delta/2$.  However, even for the
relatively small systems studied in
Figs.~\ref{fig:full_paired_towards_edge} and
\ref{fig:full_polarized_towards_edge}, the collective mode response
has a sharp onset at the value $\lambda= \lambda_\Delta/2$ at the
centre of the trap.  The appearance, in
Fig.~\ref{fig:full_polarized_towards_edge}(b), of a large response at
wavelengths slightly smaller than $\lambda_\Delta/2$ is associated
with the non-constancy of $\lambda_\Delta$, which decreases slightly
towards the edge of the FFLO phase.

We propose that the signatures of
Figs.~\ref{fig:full_paired_towards_edge}
and~\ref{fig:full_polarized_towards_edge} provide a convenient way to
detect the FFLO phase in experiment.
The sharp feature at $\lambda_\Delta/2$ appears only when the gap displays oscillatory behaviour, so is absent for a non-interacting gas and for temperatures when the FFLO phase disappears ($k_BT \simeq 0.1 E_F$ \cite{finite_temp_phase_diag_of_spin_pol_fg-liu_drummond08}).

In order to observe these
signatures in experiments one needs to be able to create a variable wavelength
optical lattice. This can be done using the technique used by
\citet{ex_spectrum_of_bec_steinhauer_davidson02}.  While the signal
appears for a spin-symmetric perturbation, the response is larger for
spin-asymmetric. Thus, any spin-dependence of the optical lattice will
enhance the ability to distinguish the peaks associated with the
oscillation of $\Delta$ from the background peaks. Optical lattices
with spin-dependence have been created by
\citet{cohenrent_transp_of_neut_atoms_in_spin_dep_opt_latt_mandel_bloch03},
and would be ideally suited to this purpose.
There are two natural ways to find the response of the spin-dipole
mode.  One way is to follow precisely the approach of the calculation,
and make the strength of the optical lattice time dependent,
as in
Ref.\cite{transitino_from_strongly_int_1d_sf_to_MI_stoferle_Kohl_Esslinger04}.
One can then selectively excite the spin-dipole mode by bringing the
temporal oscillation of the lattice in resonance with the spin-dipole
mode.  Another way (similar to the approach used in Ref.
\cite{altmeyer07}) is to apply a static periodic potential to the
system, allow it to equilibrate, and then switch off this potential
abruptly.  This will excite collective modes of many
frequencies from which the response of the spin-dipole mode can be
obtained by Fourier transform.

In either case, we emphasise that the required measurements of
particle density are on the length scale of the size of the cloud
.  Thus, our proposed method allows the
probing of the microscopic physics by perturbing the system with short
wavelengths while requiring only the measurement of densities on
the length scale of the cloud.

\acknowledgments{We acknowledge useful comments from F. Essler, C. Lobo, G. Orso and C. Salomon and
useful discussions with M. K\"ohl and N. Tammuz on the
experimental realisability. This work was supported by EPSRC.}


\begin{thebibliography}{30}
\bibitem{comment_in_fig_1}
  The response is defined as $\sqrt{\sum_i
  \left(\delta \rho_{i, \ua}^2 + \delta \rho_{i,\da}^2\right)}$. The divergent response  at
  $\omega=0$ is not shown.
\end{thebibliography}
\end{document}